\documentclass[preprint,12pt]{elsarticle}
\usepackage{amsmath}
\usepackage{graphicx}
\usepackage{titlesec}

\usepackage{txfonts}
\usepackage{caption}  % new
\usepackage{subcaption} %new
\usepackage{tabularx}
\usepackage{booktabs}
\usepackage{makecell}
\usepackage{lipsum,booktabs}
\usepackage{mathptmx}
\usepackage{amssymb}
\usepackage{mathrsfs}

\usepackage{longtable}
\usepackage{xcolor}
\usepackage{colortbl}
\usepackage{amsfonts}

\makeatletter
\def\ps@pprintTitle{%
  \let\@oddhead\@empty
  \let\@evenhead\@empty
  \def\@oddfoot{\reset@font\hfil\thepage\hfil}
  \let\@evenfoot\@oddfoot
}
\makeatother

\begin{document}

\begin{frontmatter}

\title{Investigation of the neural origin of non-Euclidean visual space and analysis of visual phenomena using information geometry}
\author [inst1]{Debasis Mazumdar}
\affiliation[inst1]{Department of computer science Sister Nivedita University={},%Department and Organization
            addressline={DG Block (Newtown), Action Area I, ½.  }, 
            city={NewTown},
            postcode={700156}, 
            state={West Bengal},
            country={India}}            
     
            \author[inst2]{Kuntal Ghosh}
\affiliation[inst2]{Machine Intelligence Unit, Indian Statistical Institute={},%Department and Organization
            addressline={203 B T Road}, 
            city={Kolkata},
            postcode={700108}, 
            state={West Bengal},
            country={India}}
%\author[inst1,inst2]{Author Three}
\author[inst3]{Soma Mitra}
\affiliation[inst3]{CDAC, Kolkata={},%Department and Organization
            addressline={Plot-E2/1, Block-GP, Sector-V, Salt Lake City,}, 
            city={Kolkata},
            postcode={700091}, 
            state={West Bengal},
            country={India}}
\author[inst3]{Late Kamales Bhaumik}
%\date{May 2025}

\begin{abstract}
%\verb|\footnotetext{Here's the footnote.}|. \footnotetext{Corresponding %e-mail:debasis.m@snuniv.ac.in footnote.}

The present paper aims to develop a mathematical model concerning the visual perception of spatial information. In vision research and cognitive science, it is usually conjectured that the processing of spatial information takes place in the mid-level vision as described by David Marr. It is a challenging problem in theoretical neuroscience to investigate how the spatial information of the objects in the physical space is encoded and decoded in the neural processes in the brain. In the past, researchers conjectured the existence of an abstract visual space where spatial information processing takes place. Based on several experimental data it was conjectured that the said psychological manifold is non-Euclidean. However, the consideration of the neural origin of the non-Euclidean character of the visual space was not explicit in the models. In the present paper, we showed that the neural mechanism and specifically the Fisher information contained in the neural population code plays the role of energy-momentum tensor to create the space-dependent metric tensor resulting in a curved space described by a curvature tensor. It is not surprising as the research in information geometry demonstrated that from the Fisher-Rao information metric, defined by the statistical distribution of any stochastic system, the Einstein tensor can be derived. The resulting tensor contains essential information about the energy-momentum tensor of a classical scalar field when the entropy data or the spectrum data of the system are embedded into the classical field as the field strength. The theoretical prediction of information geometry regarding the emergence of curved manifolds in the presence of the Fisher information is verified in the present work in the domain of neural processing of spatial information at mid-level vision. Several well-known phenomena of visual optics are analyzed using the notion of non-Euclidean visual space, the geodesics of the space, and the Fisher-Rao metric as the suitable psychometric distance.

\end{abstract}

\begin{keyword}
visual space \sep mid-level vision \sep information geometry \sep Fisher information\sep Helmholtz horopter \sep parallel and distant alleys \sep Muller-Lyer illusion.
\footnote{Corresponding e-mail:debasis.m@snuniv.ac.in}
\end{keyword}

\end{frontmatter}

%\maketitle

\section{Introduction}
\label{sec:introduction}
Considering cognitive activities as a complex system, its analysis and the development of its mathematical model is one of the active areas of research in psychophysics. Our aim in the present investigation is to develop a mathematical model of the perception of spatial attributes in human vision. More specifically the investigation is concerned with the human vision which enables us to recognize that we are surrounded by a three-dimensional manifold of objects. This manifold is designated as the physical space which is well describable by the Euclidean geometry. The objects in the physical space possess characteristic qualities like color, brightness, form, and localization. Therefore in visual perception, we are not only aware of the distribution of colors and brightness, but through highly complex cognitive processes we realize that certain of these attributes are combined to unities which we designate as objects, having definite geometrical attributes (forms) and a definite localization in a three-dimensional space. This space is designated as the visual space. Following David Marr’s computational reductionist model of the visual process \cite{marr2010vision} we conjecture that in visual perception of the spatial attributes neural mechanisms in the mid-level vision create the visual space. The challenging question is whether the visual space is metric. Can we introduce the concept of a metric in this manifold of sensation? In many perceptual experiences like the sensation of heat or brightness, we don’t feel the necessity of a metric. We merely consider whether the sensation $p_1$ is greater than, equals to, or less than the sensation $p_2$ and not concerned with how much greater or smaller. It is also not true that greater or smaller is the only attribute of the psychological manifolds that we perceive. For example in estimating the depth, the spatial separation between two objects the metric property of the space becomes evident. Historically the first prescient analysis of the problem was made by Ernst Mach \cite{mach1906space} in an influential article heading ‘On Physiological as Distinguished from Geometrical Space’. Later on Luneburg \cite{luneburg1947mathematical} published a profound mathematical model of the visual space. Luneburg’s model, though very rigorous mathematically, did not include the neural origin of the non-Euclidean nature of the visual space. It was based on the observation of the famous alley experiments conducted by Hillebrand and Blumenfeld \cite{blank1953luneburg,indow2004global}.In these experiments, it was evidenced that the visually sensed space generated from an unfamiliar spatial organization of points of light displayed in a dark background without revealing any cue (frameless scenario) may not conform to a Euclidean mapping of the physical space. Luneburg argued that to explain the experimental data of Hillebrand and Blumenfeld it is essential to consider that the visual space in binocular vision is a hyperbolic space of constant curvature. After a long silence in the research of visual space, we attempted to develop a mathematical model of the visual space considering the detailed encoding mechanisms of the spatial localization of objects and other spatial attributes by the neural population codes. The proposed model is developed in the framework of information geometry. Information geometry is a branch of information theory, wherein information is described using differential geometry \cite{amari2000methods}. The connection between physics and more specifically thermodynamics and information theory dates back to Gibbs and Boltzmann \cite{jaynes1965gibbs}. Recently the topic has been receiving attention of the researchers in different applied and basic sciences \cite{kawai2007dissipation,ito2020stochastic,kanitscheider2015measuring,ito2013information}
One of the key motivations behind the renewed interest is based on the query of whether information is a quantity, as physical as matter or energy. The same question was also raised by earlier researchers. Landauer \cite{landauer1991information,landauer1996physical}, put it up in the form of a profound question “Is information physical?” Wheeler also raised the same question in searching the links between information, and the physics of quantum era \cite{wheeler2018information}.In recent times, researchers of quantum information theory are trying to develop an information-theoretic interpretation of quantum mechanics \cite{brukner1999operationally,deutsch2000information}, while researchers in the field of information geometry explained that Fisher information can describe the emergence of general relativity \cite{matsueda2013emergent,velazquez2016riemannian}. In these studies, an information-theoretic stress-energy tensor is derived which plays the same role of mass as appeared in Einstein’s field equation of general relativity. 
In the present investigation, we describe the evidence of the physical nature of Fisher information in a completely different domain, namely the neural processes of vision. The central achievement of our investigation is to show that the Fisher information of the neural population code represented by the Gaussian probability density function plays the role of a stress-energy tensor which generates the hyperbolic nature of the visual space. The metric tensor of the visual space is calculated using the Gaussian neural population code, and using it, different geometric and topological parameters of the visual space are derived. Finally considering the Fisher-Rao distance function representing the appropriate psychometric distance of the hyperbolic visual space, several visual phenomena are modeled. The paper is organized as follows. In section 2 we represent a neuromorphic computational model to show that the tuning curve of a group of neurons, generated in response to a point source in the physical space can be modeled by Gaussian pdf. Section 3 describes the model of the visual space as a statistical parametric space. In section 4 we showed the occurrence of hyperbolic visual space under the influence of the Fisher-Information geometric stress-energy tensor and using the information geometric model of the visual space we explained several visual phenomena like the occurrence of errors in estimating the spatial distance between two point stimulus by a human observer and the Helmholtz horopters. Finally,in section 5 we conclude the paper after discussing the applications of the model for other visual optical phenomena. All the analysis and simulation is compared with the corresponding published experimental data. 

\section{Computational model of neural encoding of localization of objects in the dorsal pathway in the brain}
One of the important functions of the human visual system is to process spatial information. The processing of spatial information in the visual pathway of our brain generates control signals to guide our motor system to interact with the environment. In the past several studies have investigated the neural mechanism in the central nervous system responsible for performing this task.It is our common experience that vision is one of the major sensory inputs that provides significant spatial information. Though the other sensory signals like auditory, olfactory, haptic, etc. provide sufficient information in localizing objects in space and guiding our navigation process, we will restrict ourselves within the query, ‘how the visual information is processed within the brain and eventually performs the space perception’. The second issue that will be addressed in this section is the choice of a suitable psychometric distance function to estimate the spatial distance between two-point stimuli. The optic pathway of the visual system begins in the retina through the transduction of the input stimulus into action potentials by the rod and cone cells with the help of rhodopsin through photosensitive cycles. The transduced signal follows through several cellular circuitry in the visual pathway namely, the retinal ganglion cells (RGC), lateral geniculate nucleus (LGN), etc. to ultimately reach the primary visual cortex called the V1 in primates or area 17 \cite{dowiasch2022visual}. Visual information processing is further progressed through a network of neurons which are subdivided into several functional modules \cite{markov2013cortical}. The pioneering work of Hubel and Wiesel \cite{hubel1959receptive} explained that neurons in area V1 are orientation-selective and respond to oriented lines or gratings presented to their receptive field (RF). Therefore it can be conjectured beyond doubt that the firing of these neurons encodes orientational or angular information. Similarly, it is the generally accepted view that space perception takes place in the processing of downstream cortical areas which we designate as the mid-level vision. From area V1, visual cortical processing takes two major routes the so-called ventral and dorsal pathways \cite{ungerleider1982two}. Goodale et al.\cite{goodale1992separate} proposed their profound model explaining the distinction between object vision and spatial vision ('what' versus 'where'). They further assigned the tasks to two distinct visual stages namely the ventral and dorsal projections, responsible for object vision and spatial vision respectively. The dorsal projection, responsible for object location is a retinogeniculate projection initiated from the retina and terminated at the dorsal lateral geniculate nucleus (dLGN). Our investigation shows that the neural firing at dLGN encodes the location of objects and is analyzed after mapping in an abstract parametric statistical space called the visual space. In an influential paper Einevoll et al.\cite{einevoll2000mathematical} proposed a mathematical model of spatial transfer characteristics of dLGN using the DOG (Difference of Gaussian) model as a descriptive model for the retinal input. Following \cite{heeger1991non} they first modeled the response of a retinal ganglion cell to a visual stimulus as,
\begin{equation}%Equation 1
      \ R_g (r) = S_g\left[\iint_{r_0} G(r_0-r) s(r_0) \ dr_0 \right].      
   \end{equation}
   Here $G(r)$ is the receptive field function, and r(x,y) is the receptive field center. $s({r_0})$ represents the stimulus at ${r_0}$. The spatial integral is computed over the entire visual field.$S_g[x]$ is the transfer function of the retinal ganglion cell, taking account of the possible nonlinearities present in the cell response. Further to ensure the half-wave rectification characteristics \cite{einevoll2000mathematical} of the response of retinal ganglion cells the transfer function is defined as, $S_g[x]=xh(x)$; where $h(x)$ is the Heaviside step function defined as, $h(x<0)=0,h(x>0)=1$. It is to be noted here that the introduction of the half-wave rectification property in the transfer function $S_g [x]$ avoids negative firing rates in the model. After the seminal work of Rodieck et al. \cite{rodieck1965quantitative} the centre-surround receptive-field function for X-class retinal ganglion cells is commonly represented by the DOG function as,
   \begin{equation} % Equation 2
      \ G(x,y) = {\frac{1}{\sqrt{2\pi{\sigma_e}^2}}e^{-\frac{\left(x^2+y^2\right)}{2\sigma_e^2}}}-{\frac{w}{\sqrt{2\pi{\sigma_s}^2}}e^{-\frac{\left(x^2+y^2\right)}{2\sigma_s^2}}}.      
   \end{equation}
The DOG function is considered to be centered at $(0,0)$ and circularly symmetric. The dimensionless parameter w represents the relative strength between the center and the surroundings. If the point stimuli $s(r)$ is modeled as a Gaussian function with a vanishingly small scale factor the convolution integral in the right-hand side of Equation (1) is again a DOG function (as the convolution of two Gaussian is a Gaussian) and when clipped by the transfer function of the retinal ganglion cells it gives a positive Gaussian function to represent the population code of a group of retinal ganglion cell. To describe the process in detail Einevoll et al.\cite{einevoll2000mathematical} derived the response of the retinal ganglion cell to a circular spot of diameter d, luminance l  and having different separations between the spot center and the center of the receptive field represented by $r_g$ as given below,
\begin{equation} %Equation 3
\begin{split}
R_g(d;r_g) & = S_g\left[\textit I(L)\left.(e^{-\frac{r_g^2}{\sigma_e^2}}\sum_{m=0}^{\infty}\frac{1}{m!}{\left(\frac{r_g}{\sigma_e} \right)}^{2m}\gamma\left(m+1;\frac{d^2}{4\sigma_e^2}\right)\right.\right.\\ 
& \left.\left.-we^{-\frac{r_g^2}
     {\sigma_s^2}}\sum_{m=0}^{\infty}\frac{1}{m!}  {\left(\frac{r_g}{\sigma_s} \right)}^{2m} \gamma\left(m+1;\frac{d^2}
     {4\sigma_s^2}\right)\right)\right].  
\end{split}
 \end{equation}

 Where  $I(L) = A_1 s(L)$ is introduced as an active function and $A_1$ is the amplitude of the excitatory center. For the stimulus spot, concentric with the mean of the receptive field, the incomplete gamma function $\gamma(m,x)$ is shown to take an exponential integral form, and the expression of the response $R_g$ (d;$r_g$→0) takes the form,
 \begin{equation} % Equation 4
  R_g(d;r_g \rightarrow 0)= S_g \left[\textit{I(L)}\left({1-e^{-\frac{d^2}{4\sigma_e^2}}}-w\left(1- e^{-\frac{d^2}{4\sigma_s^2}}\right) \right) \right].   
 \end{equation}
The response corresponds to the dLGN cell perfectly aligned to the spot center and the rate of firing is maximum. Nearby cells shifted with vanishingly small separation also get excited as well and we model their response as,

 \begin{figure} % Figure 1a and 1b
     \centering
     % \begin{subfigure}[b]{0.5\textwidth}
     %     \centering
     %     \includegraphics[width=\textwidth]{Fig-1a.png}
     %     \caption{}
     %    % \label{fig:y equals x}
     % \end{subfigure}
     % \hfill
     % \begin{subfigure}[b]{0.5\textwidth}
     %     \centering
     %     \includegraphics[width=\textwidth]{Fig-1b.png}
     %     \caption{}
     %     \label{fig:three sin x}
     % \end{subfigure}     
\includegraphics[width=0.48\textwidth]{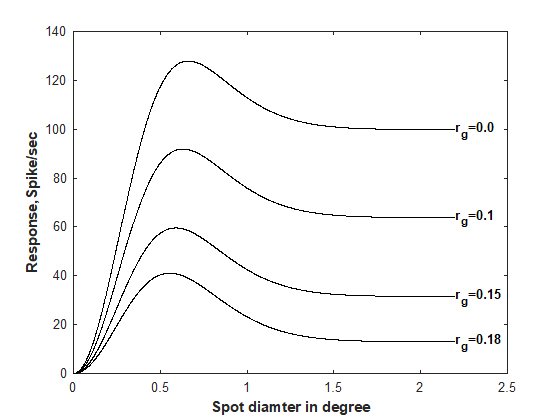}
 \includegraphics[width=0.48\textwidth]{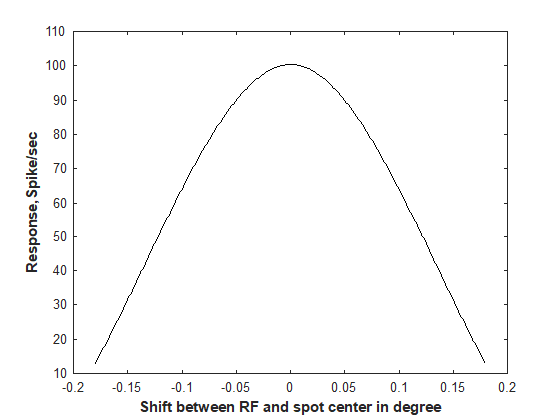}
 
        \caption{(a) Response functions of retinal ganglion cells $R_g (d;r_g )$ for circular spots of different diameters. The receptive field is considered to be a DOG with the parameters ${\sigma_e}=0.6$; ${\sigma_s} = 1.2$, $w=0.5$ and the active function $I(L)=200$ spikes/s. (b) The response of ganglion cells having a concentric receptive field with the spot center $(r_g=0)$ and cells shifted on either side of the concentric cell up to $±0.18$ degrees.}
        %\label{fig:three graphs}
\end{figure} % New ends here

 \begin{equation} % Equation 5
  R_g(d;r_g )= S_g \left[\textit{I(L)}\left(e^{-{\frac{r_g^2}{\sigma_e^2}}}\left({1-e^{-\frac{d^2}{4\sigma_e^2}}}\right)-{e^{-{\frac{r_g^2}{\sigma_s^2}}}}w\left(1- e^{-\frac{d^2}{4\sigma_s^2}}\right)\right) \right].   
 \end{equation}
 Mathematically speaking to reach Equations (4) and (5) from Equation (3) we assumed m=0. Figure 1a shows examples of retinal ganglion cell response functions both for the concentric receptive field and spot as well as for non-concentric cases. For a particular spot diameter of 2.2 degrees (as taken in \cite{einevoll2000mathematical}) the response of a group of retinal ganglion cells both concentric $(r_g=0)$ and nonconcentric $(r_g\not=0)$ up to 0.18 degrees are computed using Equations (4) and (5) as shown in Figure 1a. The response of the group of cells as presented in Figure 1b is called the population code of the neurons and is mathematically represented by the tuning curve, tuned to encode a particular attribute, the center of the point stimulus in the present case. In neuroscience, the tuning curves, after appropriate normalization, are represented by the Gaussian function \cite{pouget2000information}. The locational information of the point stimuli is encoded by the Gaussian function defined by the mean and the standard deviation respectively. The mean of the tuning curve represents the position of the endpoints. The standard deviation is significant as well because it determines the highest slope of the curve to encode the maximum change in the firing rates of nearby neurons.  For two infinitesimally close stimuli, the slope of the Gaussian tuning curve is used to discriminate them.
The output of the retinal ganglion cell further acts as the input to the dLGN cells and creates physiological couplings between the retina and dLGN and within dLGN. It is to be noted here that even though the DOG model was originally suggested for retinal ganglion cells, it has also been used by different researchers to describe receptive fields in dLGN \cite{einevoll2000mathematical}. Neurophysiological studies established that the generation of population code at the dLGN is a complicated process. The most accepted model of neural processing at dLGN can be summarized as, (i) through the relay cells they receive excitatory input from a single or a few retinal ganglion cells, (ii) they also receive feed-forward inhibition from intrageniculate interneurons which in turn receive excitation from a few retinal ganglion cells. (iii) In addition there are feedback inputs from the perigeniculate nucleus (PGN) and cortex, as well as modulatory inputs from the brain-stem reticular formation \cite{kerschensteiner2017organization}. Enviol et al. \cite{einevoll2000mathematical} considered a simplified neuronal circuit involving only the feed-forward contributions and showed that the response of the dLGN is of the same nature as the response obtained at the retinal ganglion cell. Therefore we represented the population code at the output of dLGN by appropriately normalized Gaussian function. Finally, following the established neuroanatomical description we consider that each of the thalamocortical neurons (TC) in dLGN receives input from one or few RGCs \cite{hong2011wiring,sereno2011population} maintaining separation of channels. The dorsal pathway carries the information to the anterior inferotemporal cortex (AIT) and lateral intraparietal cortex (LIP). Studying the structures associated with the ventral and dorsal visual pathways of awake-behaving monkeys and using a novel population-based statistical approach (namely, multi-dimensional scaling) it was found that a population of spatially selective LIP neurons can reconstruct stimulus locations within a low-dimensional representation with quite high precision \cite{oleksiak2010distance}.

\section{Modelling visual space as a statistical parametric space occurring at the mid-level vision and defining information geometric psychometric distance function}

Possibility of representing tuning curves as a Gaussian pdf to represent the location of a point stimulus as explained in the previous section we further extend our study to model and simulate the neural mechanism underlying the process of estimation of the spatial distance between two point stimuli. It is well known that visual information regarding the estimation of spatial distance between two point stimuli is mainly acquired either by foveating the objects through eye movement or encoding their locations through peripheral vision. Oleksiak et al.\cite{oleksiak2010distance} demonstrated the result of their psychophysical experiments to quantify the influence of encoding conditions on distance estimation. Two encoding conditions were considered namely, the fixate and the saccade visual conditions. In the fixate condition the subjects had to keep fixating at a center position in between two point stimuli marked by a cross sign and then gave their judgments of estimating the distance of separation between the two dots Figure 2a.The estimation error in each distance of separation between the dots is shown in Figure 2b. The value of absolute errors are eye estimated (from the figure presented in \cite{oleksiak2010distance}) by three human subjects and their average values are plotted in Figure 2b (in the subsequent section we will simulate the experimental result of Oleksiak et al. in the framework of our model of visual space.)\begin{figure} %Figure 2a & 2b
     \centering
     \raisebox{5mm}{\includegraphics[height=3.25cm]{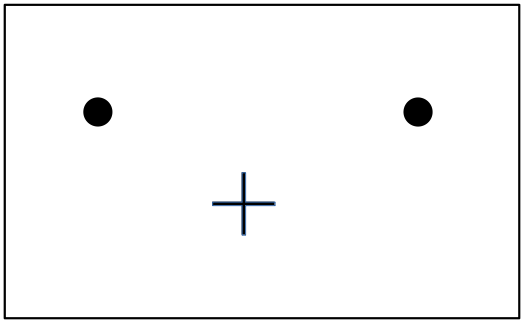}}
     \hfill
     \includegraphics[height=4cm]{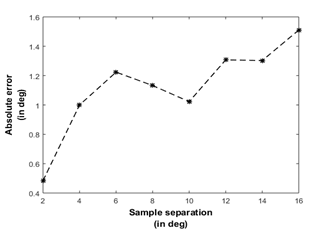}
        \caption{(a) Stimulus used in the fixate trial by Oleksiak et.al.\cite{oleksiak2010distance}. (b) Experimental data of estimation error as a function of the distance between two dots \cite{oleksiak2010distance}.}
        %\label{fig:three graphs}
\end{figure} % New ends here
 
%  \begin{figure} %Figure 2a & 2b
%      \centering
%      \begin{subfigure}[b]{0.4\textwidth}
%          \centering
%          \includegraphics[width=\textwidth]{Fig-2a.png}
%          \caption{}
%           \end{subfigure}
%      \hfill
%      \begin{subfigure}[b]{0.5\textwidth}
%          \centering
%          \includegraphics[width=\textwidth]{Fig-2b.png}
%          \caption{}
%           \end{subfigure}     
%         \caption{(a) Stimulus used in the fixate trial by Oleksiak et.al.\cite{oleksiak2010distance}. (b) Experimental data of estimation error as a function of the distance between two dots \cite{oleksiak2010distance}.}
%         %\label{fig:three graphs}
% \end{figure} % New ends here
 From the figure, one can conclude that positional uncertainty and error of localization performance increases with the distance of separation between the two dots or in other words with eccentricity. It has been shown that the positional uncertainty can be substantiated by a higher variability which ultimately reflects as an error in localization performance \cite{oleksiak2010distance,mateeff1983peripheral,toet1988effects}. We conjecture that the population code represented by the Gaussian tuning curve occurs to encode the location of a point stimulus in the human brain Fig 3.
 \begin{figure} [!ht]
     \centering
     \includegraphics[width=\hsize]{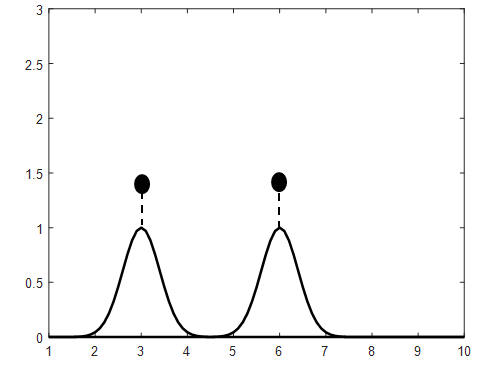}
     \caption{Encoding of two point stimulus by the Gaussian-shaped tuning curve.}
     \label{fig:enter-label}
 \end{figure}

 Further, we conjecture the existence of a statistical parametric visual space evoked at the mid-level visual processing stage on which the Gaussian population codes or the tuning curves are represented as points.  

The Gaussian population codes in the said visual space are completely defined by two parameters, namely the mean $(\mu)$ and the standard deviation $(\sigma)$ of the tuning curves. These parameters are independent of one another and hence can be plotted in a 2-D rectangular Cartesian coordinate system with axes as $\mu$ and $\sigma$. Since $\sigma > 0$, every point in the upper half of $\mu-\sigma$ plane (except on the line $\sigma = 0)$ is a Gaussian pdf. Therefore the visual space is a half-plane with frames of reference defined by the mean $\mu$ and standard deviation $\sigma$ of the tuning curve and can be written as,
 \begin{equation} % Equation 6
  \textit H_f = \left \{(\mu,\sigma)\in \mathbb{R}^2|\sigma>0 \right\}.   
 \end{equation}
 Based on these preliminary arguments we begin with modeling the visual process of estimating the distance between two-point stimuli as shown in Figure 2a. As stated earlier, in the half plane each tuning curve is represented as a point and is described by a univariate Gaussian pdf,
 \begin{equation} % Equation 7
  \int_X p\left(r,\Omega\right)dr=1.  
 \end{equation}
Where r is a stochastic variable representing the firing rate of neurons and  $\Omega=(\mu,\sigma)$ is the parameter set characterizing the tuning curve. As argued by Tkacik et al. \cite{tkavcik2013retinal} since the behavior of organisms is guided by the difference in the retinal responses therefore the biologically suitable and appropriate distance between stimuli $s_1$  and $s_2$ should also be a measure of similarity between their population codes. Following \cite{tkavcik2013retinal} we consider the Kullback-Leibler distance between the distributions, elicited by the neural responses as the retinal distance between the stimuli, and termed it as the psychometric distance. The K-L distance between two distributions $p(x;\theta)$ and  $p(x;\theta+d\theta)$ can be written as,

\begin{equation} % Equation 8
 \textit D_{kl}=\int_X p(x;\theta) ln p(x;\theta)dx - \int_X p(x;\theta) ln p(x;\theta+d\theta)dx.
\end{equation}
 This is not the usual symmetrized K-L distance for the exchange between $p(x;\theta)$ and  $p(x;\theta+d\theta)$ but its expansion up to the second order evolves into a nice expression of the  
\begin{equation} % Equation 9
\begin{split}
\textit D_{kl} & \cong -\int_X p(x;\theta)\left\{\frac{1}{p(x;\theta)}\frac{\partial p(x;\theta)}{\partial\theta^\mu}d\theta^\mu-\right.\\
&\left.\frac{1}{P^2(x;\theta)}\frac{\partial p(x;\theta)}{\partial\theta^\mu}\frac{\partial p(x;\theta)}{\partial\theta^\nu}d\theta^\mu\theta^\nu 
 \right\}dx \\
& = \frac{\partial}{\partial\theta^\mu}\left(\int_x p(x;\theta)dx\right)d\theta^\mu + \frac{1}{2} g^{\mu\nu}d\theta^\mu d\theta^\nu\\
& =\frac{1}{2} g^{\mu\nu}d\theta^\mu d\theta^\nu.
\end{split}
\end{equation}
 $g_{\mu\nu}$ is the metric tensor of the visual space, known as the Fisher-Rao information metric, and  is expressed as,
\begin{equation} % Equation 10
g_{\mu\nu}= \int_X p(x;\theta)\frac{\partial\phi}{\partial\theta^\mu} \frac{\partial\phi}{\partial\theta^\nu}dx.  
\end{equation} 
The distance between two infinitesimally separated tuning curves is represented as points $p(\mu_1,\sigma_1)$ and $p(\mu_2,\sigma_2)$ in the half plane is given by the differential line element,
\begin{equation} % Equation 11
 d^2_{sf}=\sum_{\mu\nu=1}^{2} g_{\mu\nu}d\theta^\mu d\theta^\nu. 
\end{equation}
The function  $\phi(x;\theta)=- ln p(x;\theta) $  is the negative of the log-likelihood function and is termed a spectrum.  The importance of the Fisher-Rao information metric can be realized because, in each step of measurement, we gain information about any system. The maximum amount of information that can be gained by measuring the change of any observable due to the change in the parameter is encoded in the Fisher-Rao information metric. Considering  $p(x;\theta)$ as Gaussian pdf we first compute the log-likelihood function $\phi(x,\theta)$. Finally using $\phi(x,\theta)$ different components of the metric tensor, when computed using Equation (10)  are found  as,  $g_{\mu\mu}= \frac{1}{\sigma^2}$ ,$g_{\sigma\sigma}=\frac{2}{\sigma^2}$   while $g_{\mu\sigma}=g_{\sigma\nu}=0$. Therefore the differential line element representing the proper distance in the visual space, known as the Fisher-Rao distance can be written as,   
\begin{equation} % Equation 12
 d^2_{sf}=\frac{d\mu^2+2d\sigma^2}{\sigma^2}. 
\end{equation}
When the distributions differ only in mean, the metric reduces to,
\begin{equation} % Equation 13
 d^2_{sf}=\frac{d\mu^2}{\sigma^2}.  
\end{equation}
For univariate normal distribution with variance $\sigma$ the distance between two distributions $\textit N (\mu_1,\sigma)$ and $\textit N (\mu_2,\sigma)$ is given by \cite{atkinson1981rao},
\begin{equation} % Equation 14
 s_f(\mu_1,\mu_2,\sigma)=\frac{|\mu_1-\mu_2|}{\sigma}.  
\end{equation}
After determining the fundamental metric tensor and the psychometric distance function we next represent the information geometric field equation and explain that the presence of the Fisher information causes the hyperbolic nature of the visual space. Further, we describe several topological properties of the visual space in the following section.

To find the geometric and topological properties of the visual space in the framework of information geometry, we first compute the components of the Christoffel symbol $\Gamma_{yz}^x$ as they carry significant information regarding the geometrical properties of any Riemannian space. For any n-dimensional Riemannian manifold there are $N=\frac{n^2(n+1)}{2}$  numbers of independent components of Christoffel symbols which can be computed using the equation,
\begin{equation} % Equation 15
\Gamma_{yz}^x=\frac{1}{2}g^{\beta\alpha}\left[\frac{\partial g_{\alpha\mu}}{\partial{x^\nu}}+\frac{\partial g_{\alpha\nu}}{\partial{x^\mu}}-\frac{\partial g_{\mu\nu}}{\partial{x^\alpha}}\right].  
\end{equation}
For Gaussian pdf the different components of the Christoffel symbol when calculated, we obtain, $\Gamma_{\mu\mu}^\mu=\Gamma_{\sigma\sigma }^\mu=\Gamma_{\mu\sigma}^\sigma=0$, $ \Gamma_{\mu\sigma}^\mu=\Gamma_{\sigma\sigma}^\sigma=-\frac{1}{\sigma}$  and $\Gamma_{\mu\mu}^\sigma= \frac{1}{2\sigma}$. Next, we compute the Ricci scalar curvature defined as,
\begin{equation} % Equation 16
R=g^{ij}R_{ij}.  
\end{equation}
Where $R_{ij}$ is the component of the Ricci curvature tensor. Expanding Equation 16 using the values of the component of the metric tensor $g^{ij}$ and using Einstein’s summation convention we can write the scalar curvature for the two-dimensional Fisher information space as, 
\begin{equation} % Equation 17
 R=\sigma^2 R_{\mu\sigma\mu}^\sigma+\frac{\sigma^2}{2} R_{\sigma\mu\sigma}^\mu 
\end{equation}
For Gaussian pdf, the Riemann curvature tensors are obtained as \cite{mazumdar2020representation}, $R_{\mu\sigma\mu}^\sigma=\frac{-1}{2\sigma^2}$ ,$R_{\sigma\mu\sigma}^\mu=\frac{-1}{\sigma^2}$ and finally, the scalar curvature is obtained as  $R=\sigma^2 (\frac{-1}{2\sigma^2})+\frac{\sigma^2}{2} (\frac{-1}{\sigma^2} )=-1$. Therefore we can define the visual space as a Riemannian manifold with constant negative curvature (Hyperbolic space with scalar curvature = -1) equipped with Fisher-Rao information geometric metric tensor. It is worth noting here that, in any homogeneous Riemannian space of constant curvature the form and localization are completely uncorrelated and ensures the free mobility of rigid bodies as per the Helmholtz-Lie proposition \cite{luneburg1947mathematical,indow2004global}. The experimental verification of the curvature of the visual space is done by many researchers and there are wide variations in their reported results. Blank \cite{blank1961curvature} performed an ingenious experiment using point light sources to form an isosceles triangle in the horizontal plane. In Euclidean space, the line joining the midpoints of two sides of a triangle is equal in length to half the third side while in hyperbolic space it is smaller and in spherical space, it is larger. Out of seven observers, the judgment of six observers confirms the negative sign of the curvature, i.e. hyperbolic nature of the space while one observer does not exhibit significant curvature. Hagino and Yoshioka \cite{hagino1976new} arranged light points $Q_i$ along a horizontal axis and instructed five subjects to arrange another set of light points to form circles centered around $Q_i$ s. Analyzing the result they reported that the measurement of curvature based on the data is found negative in all cases while the value of the curvature is found to vary with $Q_i$ s and the radius of the circles. These two experiments are conducted in a closed environment and can be regarded as representing frameless visual space in dark room conditions. More recently Koenderink et al. \cite{koenderink2000direct} reported their result of measurement of intrinsic curvature of the visual space based on the Gauss’s original definition, i.e. the angular excess in a triangle equals the integral curvature over the area of the triangle. The experiment was conducted in the natural environment, an open field under bright daylight conditions with everything in plain sight. The result shows that the curvature varies from elliptic in near space to hyperbolic in far space and parabolic at very large distances. Although the experimental results are not decisive and show strong dependencies on the experimental conditions and the nature of the stimuli, it can be said assertively that the frameless visual space evoked in the dark room condition is endowed with constant negative curvature. Though we do not have adequate neuro-scientific data logically it can be concluded that in the open field condition presence of different background visual cues distorts the tuning curves and ultimately the population code of neurons becomes noisy and makes it deviates from being a hyperbolic space of constant negative curvature.Next, we consider another important geometric entity, the Ricci tensor. The components of it are calculated using the equation,
\begin{equation} % Equation 18
 R_{\mu\nu}= \frac{\partial\Gamma_{\mu\nu}^k}{\partial{x^k}}-\Gamma_{\mu\beta}^k \Gamma_{\nu{k}}^\beta-\frac{\partial\Gamma_{\nu{k}}^k}{\partial{x^\mu}}+\Gamma_{\beta{k}}^k \Gamma_{\mu\nu}^\beta   
\end{equation}
The metric tensor can further be represented in an alternative way as \cite{velazquez2016riemannian},
\begin{equation} %Equation 19
 g_{\mu\nu} =-\frac{\partial^2\phi}{\partial{x^\mu}\partial{x^\nu}}+\Gamma_{\mu\nu}^k\frac{\partial\phi}{\partial{x^k}}+\frac{\partial\Gamma_{\nu{k}}^k}{\partial{x^\mu}}-\Gamma_{\beta{k}}^k \Gamma_{\mu\nu}^\beta  
\end{equation}
Combining Equations (18) and (19) we can write,
\begin{equation} %Equation 20
 R_{\mu\nu}+g_{\mu\nu}= \rho_{\mu\nu} 
\end{equation}
Where the tensor $\rho_{\mu\nu}$ is given by,
\begin{equation} % Equation 21
   \rho_{\mu\nu}= -\frac{\partial^2\phi}{\partial{x^\mu}\partial{x^\nu}}+\Gamma_{\mu\nu}^k\frac{\partial\phi}{\partial{x^k}}+ \frac{\partial\Gamma_{\mu\nu}^k}{\partial{x^k}}-\Gamma_{\mu\beta}^k \Gamma_{\nu{k}}^\beta
\end{equation}
We can further rewrite Equation (20) as,
\begin{equation} % Equation 22
 R_{\mu\nu}+g_{\mu\nu}-\frac{1}{2}g_{\mu\nu}\rho=\rho_{\mu\nu}- \frac{1}{2}g_{\mu\nu}\rho = T_{\mu\nu} 
\end{equation}
Here is the new tensor  $T_{ij}= \rho_{\mu\nu}-\frac{1}{2} g_{\mu\nu} \rho$  is introduced and  the scalar $\rho$  is obtained by tracing out $\rho_{\mu\nu}$   by the metric tensor as, $\rho=g^{\mu\nu} \rho_{\mu\nu}$.  The left-hand side of Equation (22) can be further re-written as,$R_{\mu\nu}+g_{\mu\nu}-\frac{1}{2} g_{\mu\nu} g^{\mu\nu} \rho_{\mu\nu}= R_{\mu\nu}+g_{\mu\nu}-\frac{1}{2} g_{\mu\nu} g^{\mu\nu} (R_{\mu\nu}+g_{\mu\nu} )$. In arriving at the last expression we replaced $\rho_{\mu\nu}$ using Equation (20). With this algebraic re-organization finally, we can write Equation (22) as, 
\begin{equation} % Equation 23
  R_{\mu\nu}-\frac{1}{2}g_{\mu\nu}R-\frac{1}{2}(n-2)g_{\mu\nu}=T_{\mu\nu}  
\end{equation}
Here $R=g^{\mu\nu} R_{\mu\nu}$ is the Ricci scalar, obtained by tracing out the Ricci tensor by the metric tensor and $g_{\mu\nu} g^{\mu\nu}=n$   is the dimension of the space. The tensor $T_{\mu\nu}$ resembles the stress-energy tensor. The left-hand side of Equation (23) includes the geometric entities like metric tensor, Ricci curvature tensor, and Ricci scalar of the space while the right-hand side is the stress-energy tensor, a function of Fisher information. Therefore the curvature of the space changes when information appears in it. The analysis so far indicates that the visual space appearing in the mid-level vision by the LIP neurons is a metric space and the differential distance in this space is defined by Equation (12). The appearance of the standard deviation of the tuning curve in the denominator of the distance is significant from the neuro-scientific standpoint as well. Due to several factors like the modulation of the excited region, variation of the surround, and encoding conditions like attentional state of vision, the standard deviation of the tuning curve changes, and the effect reflected in the distance measurement. Another important topological property is the compactness of the visual space. Though sequential compactness of the space of Gaussian random variables cannot be always guaranteed, one can realize a weaker convergence in which a sequence of random variables has a subsequence convergence within the distribution. With this constraint, the visual space can be considered as (weakly) sequentially compact and hence metrizable \cite{billingsley2013convergence}. Topologically any bounded infinite set of sequences in the visual space has an accumulation point \cite{indow2004global}. Extending the discussion we can further conclude that the hyperbolic lines, hyperbolic rays, and hyperbolic line segments are convex in the Euclidean map like the upper half plane or the Poincare disc \cite{anderson2006hyperbolic}. The space is differentiable too. According to Busemann, any space possessing all these properties yields geodesics \cite{indow2004global,busemann1959axioms}. The next important issue is to examine homeomorphism between different Euclidean maps of the hyperbolic visual space. The Euclidean map represented by the upper half plane model based on the Fisher-Rao distance $ds_F^2=\frac{(d\mu^2+2d\sigma^2)}{\sigma^2} =\frac{|d\omega|^2}{\sigma^2}$ , 
where $(\omega=\mu+i\sqrt[] 2\sigma)$, is homeomorphic with  the metric of the Poincare disk model,$\frac{4(d\alpha^2+d\beta^2 )}{[1-(\alpha^2+\beta^2 )]^2}  =(\frac{4|dz^2 |}{[1-|z|^2 ]})$ , where $z=\alpha+i\beta$. The latter is used in Luneburg’s model to describe different visual curves. It is easy and straightforward to verify that under the Mobius transformation $\omega = i\frac{(1+z)}{(1-z)}$   the Fisher-Rao distance is homeomorphic with the metric of the Poincare disc model apart from a constant factor of 2 \cite{mazumdar2020representation}.Naturally, all the curves derived in one model are transformable to the other conformally using the aforementioned Mobius transformation. Therefore Fisher-Rao information geometric upper half space and the Poincare disk are the two Euclidean maps of the visual space which are homeomorphic. After describing the information geometric model of the visual space and an explanation of the origin of the hyperbolic nature of the space due to the presence of information we will show in the subsequent chapters the application of the model to simulate several phenomena of visual optics.

\subsection{\textbf{Simulation of neural encoding process of estimation of  distance of separation between two point stimuli}}
As discussed in section 3, visual information about object locations in the nearby environment is acquired through two visual conditions either by bringing these items onto the fovea with an eye movement or by encoding their presence by peripheral vision. Different estimation biases occur in the two encoding conditions. Oleksiak et al. \cite{oleksiak2010distance}conducted psychophysical experiments to quantify the biases occurring in these two visual conditions. Our investigation is focused on the simulation of biases occurring in the estimation of the distance between two dots in fixate conditions using the model of the underlying neural mechanism and the properties of the hyperbolic visual space. The choice of fixate visual condition widens the scope of explaining several visual phenomena like the Helmholtz horopters, the parallel and distance alleys, and the Muller Lyer illusion in a common framework of hyperbolic visual space. This is important since a model is of less value if it is limited in explaining a particular experimental data rather than demonstrating its applicability in a different set of experimental data.The simulation starts with retinal mapping of the two spots stimuli used in the experiment of Oleksiak et al. \cite{oleksiak2010distance}. All the parameters used in the simulation are kept identical to those considered by Oleksiak et al. in their experiment. The visual process of projection of two dots in physical space (shown in Figure 2a) on the retina when viewed by a subject fixing the vision at the cross mark can be schematically represented as shown in Figure 4. The principal axis of the lens in the human eye is oo', ’a’ is the position of one target spot of diameter 0 .1° of visual angle at a distance oc = 65 cm from the subject. The projected spot on the retina is formed at ‘b’ ( actually in optical chiasm it gets crossed and projected on the opposite side of the principal axis, but considering the symmetry of the two dots we consider a simplified diagram that preserves the geometrical relations). The distance between the lens and the retina is taken as co' = 2.0 cm. Computer-simulated retinal projections of the two dots are plotted using a scale factor of 1 unit = 40 pixels to make the graphical presentation interpretable. The computation of different geometric parameters is explained in the caption of Figure 4. The simulated retinal projection of two dots is shown in Figure 5a. Figure 5b represents the center-surround RF profile of the retinal ganglion cell represented by the DOG function. The generated response of the retinal ganglion cell obtained after convolving the DOG RF profile with the retinal two-dot image is represented in Figure 5c. Figure 5d. represents the response of the ganglion cell after half-wave rectification by the retinal ganglion cell transfer function $S_g (x)$. As already explained above, the two Gaussian functions occurring after half-wave rectification of the response of the retinal ganglion cell is the tuning curve of the population code at the level of the retinal ganglion cell. The location of the two dots presented in the physical space is encoded as the peak of the tuning curve and they progress through the dorsal pathway maintaining the same Gaussian shape to reach the lateral intraparietal cortex where they appear as points in an abstract hyperbolic visual space as modeled above and the distance between them can be computed using Equation (14). Computation of the distance between two dots, therefore, follows the following steps:
\begin{figure} % Figure 4
   \centering
   \includegraphics[width=\hsize]{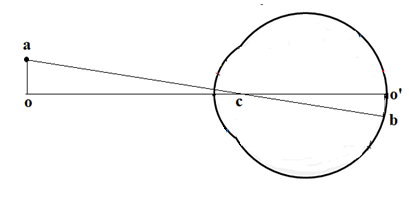}  
        \caption{Ray diagram of projection of a point stimulus at point ‘a’ in the physical space onto
 the retina at point ‘b’. The distance of the stimulus from the subject is $oc=65 cm$. The distance between the lens of the subject’s eye to the retina is taken to be  $co'= 2.0$ cm. Using simple trigonometric relation $(ao=oc \tan(\angle{aoc})$ we calculated the linear distance of the dot and their linear separation is computed as 2 × ao. The linear separation of the two dots projected on the retina is computed in the same way.}
   \label{FigVibStab}
   \end{figure}
   
Step 1: Generation of DOG function representing the center-surround RF of the retinal ganglion cell after appropriate selection of the scale factors of the excitatory $(\sigma_e)$ and the inhibitory $(\sigma_s)$ Gaussians.The variation of the classical receptive field (excitatory region) size in the retinal ganglion cell with the eccentricity of the object has been studied vastly by visuotopic mapping \cite{hubel1974uniformity,polimeni2006multi}. Vast research results provide a mathematical description of the linear relationship between the RF size and the eccentricity of the object. Following Oleksiak et al, \cite{oleksiak2010distance} we used the coarse but simple description of the scaling of the RF size with eccentricity \cite{yazdanbakhsh2008new} as given by,

\begin{equation} % Equation 24
 \sigma_e=\frac{E_{cc}^o+\alpha}{\kappa}   
\end{equation}
Where $\alpha=0.7$ and $\kappa =15$ are two constants. Since in fixating vision, the two dots appear symmetrically the eccentricity $E_{cc}^o$ is taken as half of the angular separation of the dots. The value of the scale factor of the suppression region is computed using the relationship $\sigma_s=1.6\sigma_e$  as determined by Marr et al. \cite{marr1980theory}.
Step 2: Generated DOG function is convolved with the retinal image of the two dot stimulus which generates the response of the retinal ganglion cells in the form of DOG with their center at the position of each dot as shown in Figure 5c.
Step 3: The positive going center of the response function of retinal ganglion cells is clipped using the ganglion cell transfer function $S_g (x)$.
Step 4: For two dot stimuli the generated Gaussian tuning curves are identical having different means but the same standard deviation (for a particular distance of separation) and assumed that they progress keeping the shape unchanged through the dLGN to appear as the points on the parametric visual space possibly occurring at the anterior inferotemporal cortex (AIT) and lateral intraparietal cortex (LIP). At this stage, the mean and standard deviations are computed for the two Gaussian population codes and the distance between them is computed using Equation (14). 
Step 5: The absolute error in the estimation of the distance of separation between the two dots is computed as, Error= $|L_0-L_e |$. Where $L_0$ is the actual separation between the two dots and $L_e$ is the distance of separation computed in the previous steps. 
The error computed is plotted against the separation of the two dots (in deg) and it is shown by the dotted curve in Figure 6a. The curve obtained by plotting the experimental values obtained by Oleksiak et al. \cite{oleksiak2010distance} is shown by the -* line. Comparing the two results we can state that up to 4° the two curves are closely matched but beyond that, the simulated result differs from the experimental value and the difference is increasing as the distance of separation increases. To address the problem we note the following issues described by the earlier researchers. Oleksiak et al. \cite{oleksiak2010distance} have shown that the separation by eccentricity ratio in ‘fixate’ trials changes from less than 1 to larger than 1 as a function of the presented distance. Therefore as the distance of separation between the two dots increases the eccentricity of their projected image does not change proportionately. The second important factor involved is a top-down mechanism using selective attention to allocate neuronal processing resources to the most relevant subset of the information provided by the sensors of the organism. Attentional selection of a spatial location modulates the spatial-tuning characteristics. The effect of the tuning changes includes a shift of receptive field centers toward the focus of attention and a narrowing of the receptive field when the attentional focus is directed into the receptive field. Attentional modulation of neural response based on spatial feature selection is 
 \begin{figure}[!ht] % Figure 5a,5b,5c and 5d
     \centering
     \hspace{1cm}
     \begin{subfigure}[b]{0.2\textwidth}
         \centering
         \includegraphics[width=\textwidth]{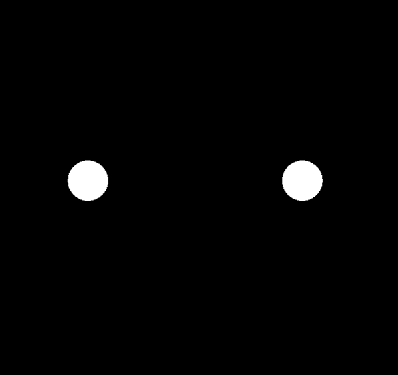}
         \caption{}
        % \label{fig:y equals x}
     \end{subfigure}
     \hfill
     \begin{subfigure}[b]{0.4\textwidth}
         \centering
         \includegraphics[width=\textwidth]{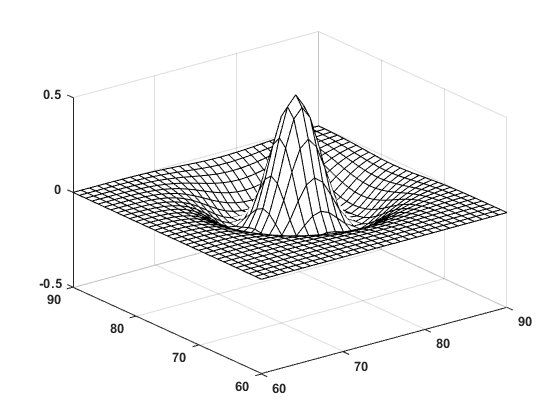}
         \caption{}
             \end{subfigure} 
     \hfill
     \begin{subfigure}[b]{0.4\textwidth}
         \centering
         \includegraphics[width=\textwidth]{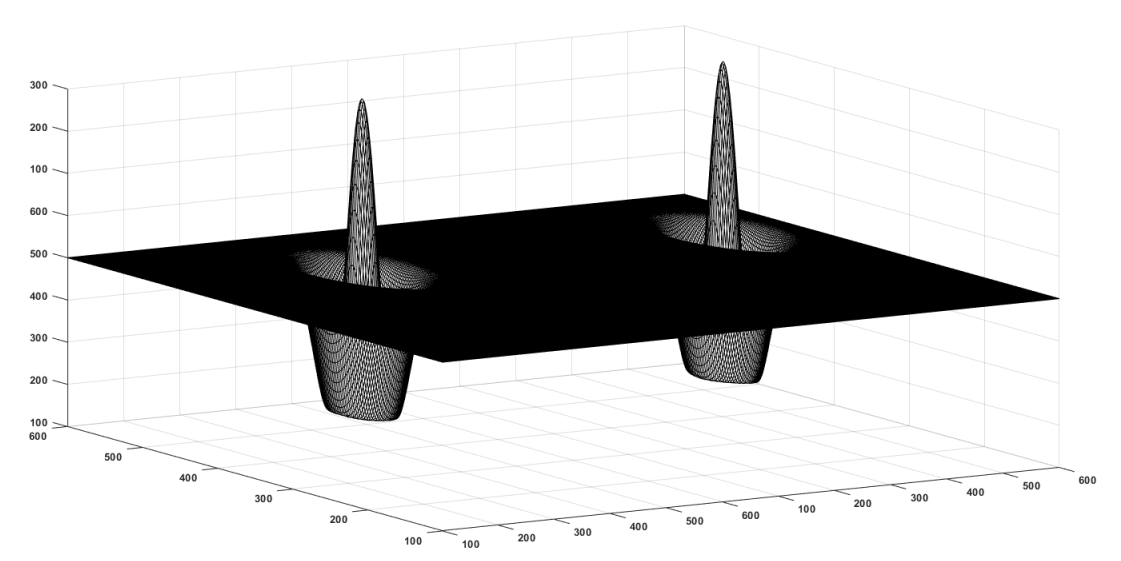}
         \caption{}
              \end{subfigure} 
     \hfill
     \begin{subfigure}[b]{0.4\textwidth}
         \centering
         \includegraphics[width=\textwidth]{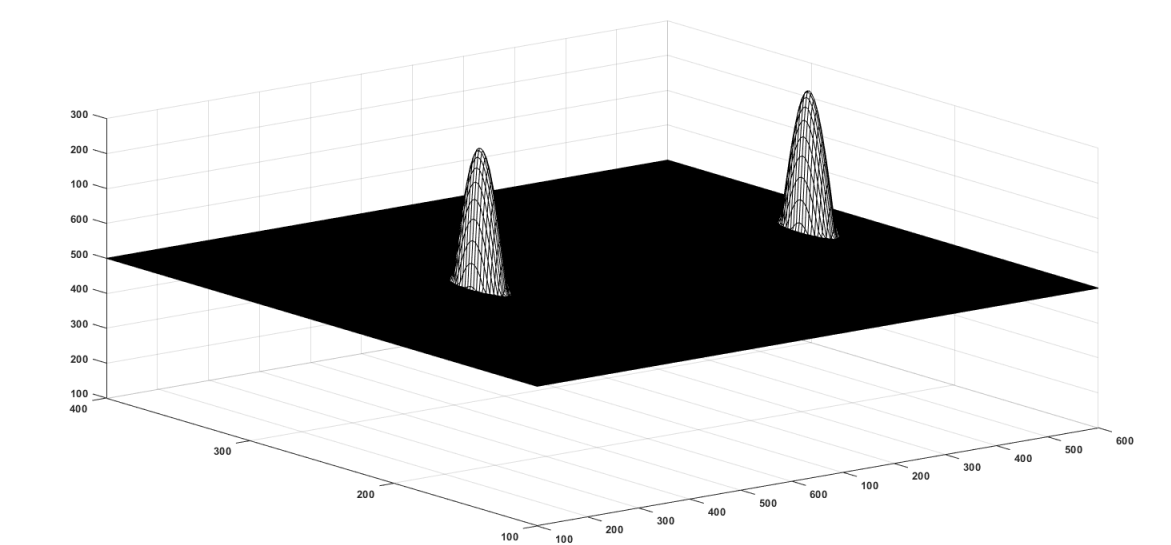}
         \caption{}
              \end{subfigure} 
        \caption{(a) computer-generated retinal projection of two dots. (b) The center surrounds the RF profile of the retinal ganglion cell. (c) Response profile of retinal ganglion cell for the two dot stimulus. (d) Half wave rectified response profile of the ganglion cell for two dot stimulus.}
        %\label{fig:three graphs}
\end{figure} % New ends here
a well-researched problem \cite{Treue,JC.Martınez-Trujillo,Bichot}. Based on these research outcomes we first modify the algorithm such that the initial value of $\sigma_e$ is set using Equation (24). Then we keep on modifying the value of $\sigma_e$ iteratively till the difference between the simulated and experimental values of the error reaches a minimum. The simulated curve is shown by the curve presented by the (–o) line and the error computed using Euclidean distance is shown by the solid line in Figure 6b. The correlation between the experimental (-*) and simulated curve (-o) is computed and found to be 0.9224 while the curve obtained using Euclidean distance and plotted by the solid line is completely in disagreement with the experimental values. The corresponding variation of $\sigma_e$ is shown in Figure 6c. It is observed from the graph presented in Figure 6c that up to 4° of separation of the dots, $\sigma_e$ increases linearly and beyond that it decreases. Such behavior of $\sigma_e$ clearly indicates that beyond 4° separation, additional effects like the attentional selection contribute significantly to modulating the neural response. Further to note that the interaction of attentional influence and the neural response has two contrastive natures, the second important factor involved is a top-down mechanism using selective attention to allocate neuronal processing resources to the most relevant subset of the information provided by the sensors of the organism.Further to note that the interaction of attentional influence and the neural response has two contrastive natures, namely the multiplicative and non-multiplicative \cite{Treue}. Multiplicative influences are dominant for tuning the features like orientation or direction of motion of objects into the neurons in the same visual areas \cite{Treue,JC.Martınez-Trujillo,Bichot}. On the other hand, non-multiplicative attentional effects are particularly predominant and reflect the influence of attention on the spatial-tuning characteristics of neurons in the extrastriate visual cortex \cite{McAdams,Connor}, as originally predicted by Moran and Desimone \cite{Moran}. On the face of such contrasting effect of attention, Womelsdorf et al. \cite{Womelsdorf,Womelsdorf1} demonstrated a general attentional gain model in which they showed that non-multiplicative attentional modulations of basic neuronal-tuning characteristics could be a direct consequence of a spatially distributed multiplicative interaction of a bell-shaped attentional spotlight with the spatially fined-grained sensory inputs of MT neurons. They used a bell-shaped spatial attentional spotlight providing multiplicative spatial weighting on the sensory inputs of MT neurons. The use of a Gaussian bell-shaped function ensured the fact that the result of a multiplicative interaction of two Gaussians still is a Gaussian. In the proposed model sensory feed-forward spatial inputs are modelled with a Gaussian connectivity profile, having a scale factor $\sigma_e$ and the spatial attention profile is also modeled by a bell-shaped function of scale factor $\sigma_A$ (say). The effect of attention is conjectured to be multiplicative generating a resultant attended response profile which is again a Gaussian with an effective scale factor $\sigma_R$ given by,

\begin{figure}[!ht] % Figure 6a,6b and 6c
     \centering
     \begin{subfigure}[b]{0.3\textwidth}
         \centering
         \includegraphics[width=\textwidth]{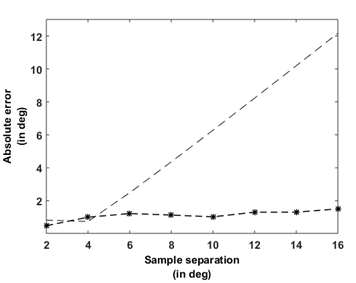}
         \caption{}
        % \label{fig:y equals x}
     \end{subfigure}
     \hfill
     \begin{subfigure}[b]{0.3\textwidth}
         \centering
         \includegraphics[width=\textwidth]{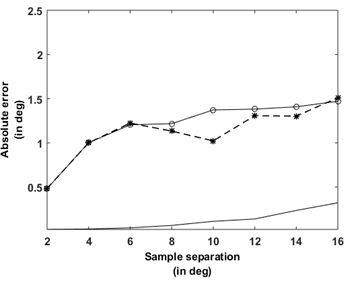}
         \caption{}
             \end{subfigure} 
     \hfill
     \begin{subfigure}[b]{0.3\textwidth}
         \centering
         \includegraphics[width=\textwidth]{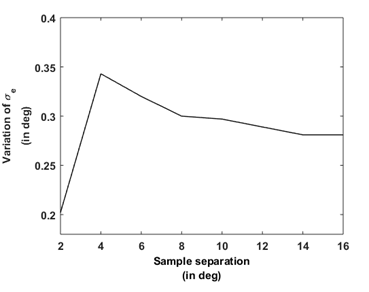}
         \caption{}
              \end{subfigure} 
            \caption{Simulation of error in distance estimation. (a) Experimental values obtained in [7] are represented by the curve (-*) simulated values of the error computed using $\sigma_e$  as given by Equation (20)  is represented by the (--) curve. (b) The simulated values of the error computed using iteratively adjusted $\sigma_e$ is represented by curve (--o) experimental data are plotted by curve (--*) and the error computed using Euclidean distance between the peaks of the Gaussian tuning curves is plotted using a solid line. (c) Variation of  $\sigma_e$  to obtain the closest match of the simulated values of the error with the same obtained experimentally. }
        %\label{fig:three graphs}
\end{figure} % New ends here
\begin{equation} %Equation 25
 \frac{1}{\sigma_R^2} =\frac{1}{\sigma_e^2} + \frac{1}{\sigma_A^2} 
\end{equation}
We considered $\sigma_A$ is the scale factor of the Gaussian neural response of the ganglion cell at a
particular eccentricity and computed using Equation (24) while $\sigma_R$  is the value of the scale factor offering the best matches of error of estimation of distance with the corresponding experimental values as given in Figure 6c. When we solved Equation (25) for $\sigma_e$ for a given $\sigma_A$,we obtain the required values of the scale factor of the bell-shaped spatial attentional spotlight as shown in Figure 7a. The corresponding attentional gain given by the amplitude of the Gaussian and given by $A_{att}= (2\pi\sigma_R^2)^{-\frac{1}{2}}$. The increase of attentional gain with the distance of separation is presented in Figure 7b. With this analysis, we finally modified the algorithm in the following way. In step 1 we initialized the value of $\sigma_A$ using Equation (24) and initializing the value of $\sigma_e$ with an initial guess. With these two initial values, we computed $\sigma_R$ using Equation (25). Thereafter estimated length is computed following steps 2 to 4. Finally the absolute error in the estimation of the distance of separation between the two dots $(L_e)$ are computed using step 5. 
The process iteratively adjusted the value of $\sigma_e$ and updated the value of $\sigma_R$  until Error=$|L_0-L_e|$ attains a minimum. The simulation described above reveals the important fact underlying the visual process of distance estimation. It advocates in favor of the existence of hyperbolic visual space and the non-Euclidean Fisher-Rao distance as the appropriate distance function to represent the psychometric distance. It is therefore natural to use them in explaining other visual phenomena.In the following, we shall derive the equation of the geodesics in the visual space and will describe that the Helmholtz horopters are the geodesics of the hyperbolic visual space.
\begin{figure}[!ht] % Figure 7a and 7b
     \centering
     \begin{subfigure}[b]{0.3\textwidth}
         \centering
         \includegraphics[width=\textwidth]{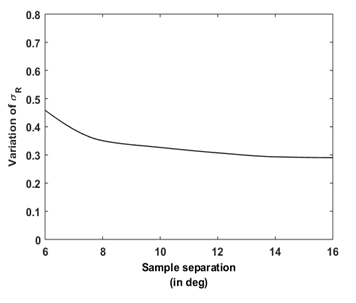}
         \caption{}
        % \label{fig:y equals x}
     \end{subfigure}
     \hspace{1cm}
     \begin{subfigure}[b]{0.3\textwidth}
         \centering
         \includegraphics[width=\textwidth]{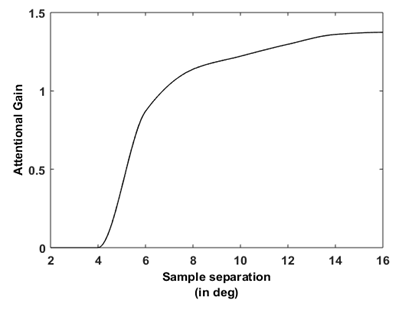}
         \caption{}
             \end{subfigure} 
    
            \caption{(a)Variation of the scale factor of the bell-shaped spatial attentional spotlight $\sigma_e$ as a function of separation between two dots in the range of in the range of 6° to 16°, assuming that the effect of attention is predominant in this range (b) Variation of attentional gain in the same range.}
        %\label{fig:three graphs}
\end{figure}

\subsection{\textbf {Simulation of the Helmholtz horopters as the geodesics of the hyperbolic visual space}}
Horopters are the locus of points in space wherein for a given fixation of the eyes the object will be projected upon the corresponding retinal points. The study of horopters remained a challenging research topic in psychology and physiological optics concerning binocular vision, for a long period. The analysis of the horopters is important as it provides information regarding the perception of direction and depth. Helmholtz pursued thorough research on horopters \cite{Helmholtz,Zajaczkowska} and described a set of lines as the horopters (later named after him) which are not always physically straight. The form and the subjective feeling of the straightness of the curves depend on the distance between the curves and the observer. To provide a brief description of Helmholtz’s horopter we consider the figure 8.
\begin{figure}[!ht] 
\centering
  \includegraphics[width=0.3\textwidth]{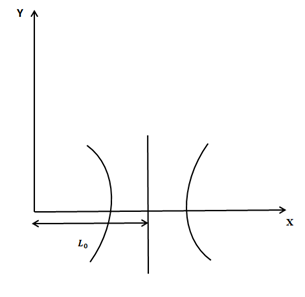}
   \caption{Helmholtz’s horopter.}
           \end{figure}
In the figure, the intersection of the horizontal plane at the eye level and the median plane of the observer is represented by the x-axis. The y-axis on the other hand is the intersection between the same horizontal plane and the frontoparallel plane passing through the two eyes (L and R). The horizontal lines that appear subjectively as parallel to the y-axis are not always parallel in the physical space. The shape of the perceived frontoparallel lines when mapped in the physical space depends on the distance between the observer and the frontoparallel lines on the x-axis. More specifically the form appears to be physically parallel lines at a certain distance. When the observer is at a nearer distance it becomes concave and convex at a further distance.  In a past publication \cite{mazumdar2020representation} we derived the locus of the geodesics of the hyperbolic visual space by solving the geodesic equations;
\begin{subequations} % Equation 26(a) and 26(b) 
\begin{equation}
 \frac{d^2\mu}{d{s^2}}+\left[\Gamma_{\mu\mu}^\mu \frac{d\mu}{ds}\frac{d\mu}{ds} + \Gamma_{\mu\sigma}^\mu \frac{d\mu}{ds}\frac{d\sigma}{ds}+ \Gamma_{\sigma\mu}^\mu \frac{d\sigma}{ds}\frac{d\mu}{ds}+\Gamma_{\sigma\sigma}^\mu \frac{d\sigma}{ds}\frac{d\sigma}{ds}  \right] =0; 
\end{equation} 
\begin{equation}
 \frac{d^2\sigma}{d{s^2}}+\left[\Gamma_{\mu\mu}^\sigma \frac{d\mu}{ds}\frac{d\mu}{ds} + \Gamma_{\mu\sigma}^\sigma \frac{d\mu}{ds}\frac{d\sigma}{ds}+ \Gamma_{\sigma\mu}^\sigma \frac{d\sigma}{ds}\frac{d\mu}{ds}+\Gamma_{\sigma\sigma}^\sigma \frac{d\sigma}{ds}\frac{d\sigma}{ds}  \right] =0.    
\end{equation}
\end{subequations}
The general equation of the geodesics in the visual space represented by the upper half plane model is obtained as \cite{mazumdar2020representation},
\begin{equation} % Equation 27
  2\sigma^2 + \mu^2-C\mu=1.  
\end{equation}
Also, in the previous section, we demonstrated that the Fisher-Rao distance function defined in the upper half plane model is homeomorphic to the distance function defined in the Poincare unit disc model under a Mobius transformation. Naturally, all the curves derived in one model are transformable to the other conformally. Therefore, we obtain the equation of the Helmholtz horopters in the Poincare disc model using the same Mobius transformation\cite{mazumdar2020representation} as given below,
\begin{equation} % Equation 28
  \frac{K_3}{4} (\alpha^2+\beta^2) -1 = C\alpha. 
\end{equation}

\begin{figure}[!ht] % Figure 9a and 9b
     \centering
     \begin{subfigure}[b]{0.3\textwidth}
         \centering
         \includegraphics[width=\textwidth]{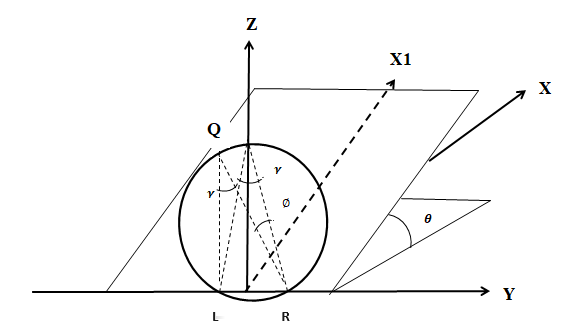}
         \caption{}
        % \label{fig:y equals x}
     \end{subfigure}
     \hspace{2cm}
     \begin{subfigure}[b]{0.3\textwidth}
         \centering
         \includegraphics[width=\textwidth]{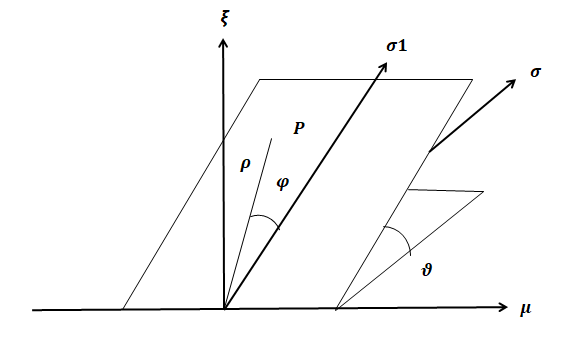}
         \caption{}
             \end{subfigure} 
    
            \caption{Coordinates of the physical and the visual space. (a) The physical space is represented by a Cartesian and a bipolar coordinate system. The circle crossing through left (L) and the right (R) is called the Vieth-Muller circle or the basic circle. Any point on the Vieth-Muller circle projects its image on the corresponding points of the two retinae. X1 is the projection of the x-axis in the plane of observation. (b) The visual space is represented by a Cartesian coordinate system and a polar coordinate system. $\sigma_1$ is the projection of the $\sigma$ axis on the plane of observation. In the physical space Q is the object and the corresponding percept in the visual space is represented by P. Other parameters are explained in the text.}
        %\label{fig:three graphs}
\end{figure}
Where $K_3$ and C are two constants. It is to be noted here that Equation (28) is the same as the locus of the Helmholtz horopters described in Luneburg’s model \cite{luneburg1947mathematical}. To transform the coordinates of the points on the geodesics in the hyperbolic visual space to the physical space we use the transformation equations proposed by Luneburg \cite{luneburg1947mathematical} and detailed in \cite{mazumdar2020representation}. The Luneburg transformation describes the physical space using the Cartesian coordinate (X, Y, Z) and the bipolar coordinates $(\gamma,\Phi,\Theta)$ as shown in Figure 9a. The coordinate $\gamma$ is called the bipolar parallax and it varies from 0 to $\pi$ in degree, depending on the distance of the object from the subject. The angular coordinate $\Phi$ and $\Theta$ are called the bipolar latitude and the angle of elevation respectively. The use of bipolar coordinates is more appropriate in describing the perceptual factors of binocular vision. For example, binocular vision besides providing depth perception gives information with a greater certainty regarding the directional localization, namely orientation relative to the baseline of the two eyes. It is easier to model this using a bipolar coordinate system.

\begin{figure}[!ht] % Figure 10a,10b and 10c
     \centering
     \begin{subfigure}[b]{0.3\textwidth}
         \centering
         \includegraphics[width=\textwidth]{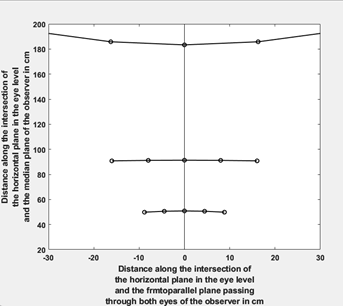}
         \caption{}
        % \label{fig:y equals x}
     \end{subfigure}
     \hfill
     \begin{subfigure}[b]{0.3\textwidth}
         \centering
         \includegraphics[width=\textwidth]{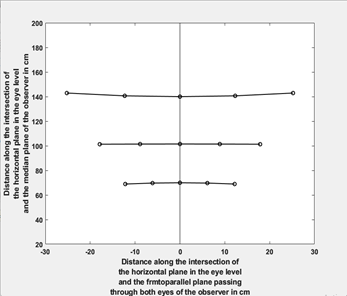}
         \caption{}
             \end{subfigure} 
    \hfill
     \begin{subfigure}[b]{0.3\textwidth}
         \centering
         \includegraphics[width=\textwidth]{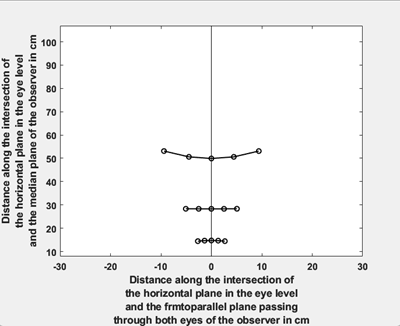}
         \caption{}
             \end{subfigure} 
            \caption{Simulated frontal plane horopters. Experimental data for three subjects are collected from \cite{Zajaczkowska} and represented by the circles. The simulated horopters are represented by solid lines. All the experimentally measured values of the personal constants used in the simulation are collected from \cite{Zajaczkowska} and presented in Table-1.}
        %\label{fig:three graphs}
\end{figure}
Luneburg further introduced the Cartesian and the corresponding polar coordinate to describe the visual space. The Cartesian coordinate of the visual space is defined by the three coordinates $(\mu,\sigma,z)$ and the polar coordinates are defined by $(\rho,\phi,\theta)$. Finally the relations between the physical and the visual space is proposed by Luneburg as,
\begin{subequations} %Equations 29(a),29(b) and 29(c)
    \begin{equation}
      \rho= 2 e^{-s\gamma},  
    \end{equation}
    \begin{equation}
        \Phi=\phi,
    \end{equation}
    \begin{equation}
         \Theta=\theta.
    \end{equation}
\end{subequations}
It is to be noted here that the distance between the centres of the two eyes (L and R in Figure 9a) is taken as 2 units and s is a personal constant of the observer, purely psychological in nature and related to the perception of depth. Using these transformation equations it is straightforward to derive the loci of the Helmholtz horopter in the visual space.
%insertion of table
\begin{table}
    \centering
    \begin{tabular}{|c|c|c|c|} \hline    
        Observer & $\sigma$ &\thead{Inter-pupillary distance \\$\nu$ in Cm} & \thead{Experimentally \\measured value\\of $\textit L_0$ in Cm} \\ \hline 
        Observer1 & 12.32 & 6.70 &101.51 \\ \hline 
        Observer2 & 12.32 & 6.48 & 89.70\\ \hline 
        Observer3 & 3.32 & 6.60 & 28.46\\ \hline
    \end{tabular}
    \caption{Experimental data \cite{Zajaczkowska} of personal constants used in the simulation of the Helmholtz horopter}
    \label{tab:my_label}
\end{table}
%End of Table insertion

Following \cite{indow2004global} equations relating the coordinates of the visual space and that of the physical space are obtained as described below,
\begin{equation} % Equation 30
    \tan\left(\frac{\gamma_0}{2} \right)\tanh[s(\gamma_0+\kappa)]-\frac{1}{4\sigma}=0,
\end{equation}
and
\begin{equation} %Equation 31
    \textit L_0= \frac{\nu}{\gamma_0}
\end{equation}
Where $L_0$ is the distance of the intersection of the straight line horopter with the x-axis, $\nu$ is the inter-pupillary distance of the subject, $\gamma_0$ is the bipolar parallax on the x-axis and $\kappa= \frac{\ln\left[\frac{1}{-K_3}\right]}{2\sigma}$
Solving the transcendental Equation (30) one can compute the value of $\gamma_0$ and with the aid of Equation (31), the distance of the straight line horopter $\textit L_0$  on the x-axis can be computed. Further, we can obtain the x and y coordinates corresponding to the points on the geodesics in the visual space using the equations given below \cite{indow2004global}, 
\begin{subequations} % Equation 32(a) and 32(b)
    \begin{equation}
        x=\frac{\nu}{2}\frac{\cos(2\phi)+\cos(2\sigma)}{\sin(\sigma)}
    \end{equation}
    \begin{equation}
        y=\frac{\nu}{2}\frac{\sin(2\phi)}{\sin(\sigma)}
    \end{equation}
\end{subequations}
We simulated the Helmholtz horopter curves for three subjects whose personal constants were experimentally measured by Zajaczkowska et al. \cite{Zajaczkowska}. The results are given in Figure 10.
\section{Conclusion}
The present paper addresses an important issue of the neural origin of visual space. Geometrical models of visual space as a Riemannian space of constant negative curvature were developed by early researchers and explanations of several unusual visual phenomena like Ames distorted rooms, Helmholtz horopters, and Parallel and distance alleys were explained successfully. However, the past models were lacking in reconciling the mathematical models with the neural mechanisms. At this juncture, the present paper proposes an alternative model of the visual space and explains its possible neural origin. Based on the results of information geometry it is shown that the presence of Fisher information in the neural population codes changes the curvature of the visual space. Therefore the physical nature of information is verified in the investigation in the domain of information processing in the brain. Also, the Fisher-Rao information metric is found to be the most appropriate metric of the psychological manifold. A number of visual phenomena involving spatial information processing are modeled and simulated in the framework of non-Euclidean visual space and using the Fisher-Rao metric. All the simulations are compared with published experimental data. In several past publications, we simulated successfully the illusory effects occurring in the case of parallel and distant alleys \cite{mazumdar2020representation} and the famous Muller-Lyer illusion \cite{Rahul,Debasis}. It will not be unreasonable to predict the applicability of the model in analyzing the illusory effects occurring in similar geometrical optical illusions named after Zollner, Hering, and Poggendorff. The proposed model can also be extended to model many real-life applications of fundamental and engineering importance. For instance the specific neurons, namely the place cells which are responsible for the perception of coordinates of the objects in physical space are one of the influential discoveries of O’keefe et al. \cite{O’keefe}. Studying the place units in the hippocampus of the freely moving rat they concluded that the hippocampal formation in both rats and humans is involved in spatial navigation. The present model which is developed for stationary stimulus can be extended to include temporal components and can be used to develop neuromorphic algorithms for robotic navigation or similar problems of computer vision. Finally, the authors have followed the COPE guidelines for ethical responsibilities. There is no potential conflict of interest.

\bibliography{sample}
\bibliographystyle{unsrt}

\end{document}